\documentclass[prl,twocolumn,floatfix,showpacs]{revtex4}
\usepackage{amssymb}
\usepackage{amsmath}
\usepackage{epsfig}
\usepackage{graphicx}

\begin{document}

\title{Symmetry background of the fractional Aharonov-Bohm oscillation
and an oscillation in dipole-transitions of narrow quantum rings
with a few-electrons}

\author {C. G. Bao*}
\author {G. M. Huang}
\author {Y. M. Liu}

\affiliation{The State Key Laboratory of Optoelectronic Materials
and Technologies, and Department of Physics, Zhongshan University,
\ Guangzhou, 510275, P.R. China}

\begin{abstract}
The low-lying spectrum of a 3-electron narrow ring has been
analyzed analytically. \ A phase-diagram for the ground state band
against the magnetic field and the radius of the ring is obtained.
The symmetry background \ of the fractional \ Aharonov-Bohm
oscillation has been revealed. A very strong oscillation in the
dipole transition is found. \ The discussion can be generalized to
N-electron rings.
\end{abstract}

\pacs {73.23.Ra,  78.66. -w\\
       *Corresponding Author.} \maketitle

\vspace{1pt}

Quantum rings with a radius from 20 to 120 nm containing only a
few electrons can now be fabricated in laboratories$^{1,2}.$ \ \
When a magnetic field $B$\ is applied, interesting physical
phenomena, e.g., persistent currents \ and Aharonov-Bohm
oscillation (ABO) with a period  $\Phi _{0}=hc/e$ ,the flux
quanta, have been observed $^{3,4}$. \ In particular, the
fractional ABO (FABO) with a period a fraction of $\Phi _{0}$ has
been observed recently$^{2,5}$. \ \ The experimental observations,
including the fractional periodicity, can in general be explained
theoretically. \ Energy spectra and electronic correlations have
been studied in quite detail$^{6-11} $.   \ Nonetheless, in the
previous study the emphasis was played in the dynamical aspect,
the effect of symmetry was less touched. \ However, this effect is
believed  to be in general important to few-body systems$^{12-16}.
$

\qquad To remedy, the motivation of this paper is to clarify how the
internal structures and physical properties are affected by symmetry. \ For
this purpose, we shall analyze the nodal structures of  wave functions,
thereby to clarify the symmetry background of the FABO and\  to clarify the
regularity appearing in the spectra. Furthermore, since the energy levels
are affected by symmetry, the optical properties might be affected as well,
this is also a point to be studied.\ \ \ In this paper only narrow rings are
concerned, they are treated as one-dimensional.

\qquad Let a magnetic field $B$\ be perpendicular to the plane of
a ring containing $N$ electrons, let $\theta _{j}$ be the
azimuthal angle of the $j-th$ electron, let a set
of relative angles be defined as $\varphi _{1}=\theta _{2}-\theta _{1},$ $%
\;\varphi _{2}=\theta _{3}-(\theta _{2}+\theta _{1})/2,\cdot \cdot
\cdot \cdot \cdot ,$ $\varphi _{N-1}=\theta
_{N}-\sum_{i=1}^{N-1}\theta _{i}/(N-1), $ and $\theta
_{C}=\sum_{i=1}^{N}\theta _{i}/N$. \ \ The Hamiltonian reads

\qquad \qquad $H=H_{coll}+H_{in}+H_{Zeeman}\qquad \qquad (1)$

where

\vspace{1pt}\qquad \qquad \qquad $H_{coll}=D(-i\frac{\partial }{\partial
\theta _{C}}+\alpha B)^{2}\qquad \qquad (2)$

is for the collective motion, \ $D=\frac{\hbar ^{2}}{2N\;m^{\ast }R^{2}}$, \
$\alpha =\frac{N\;eR^{2}}{2\hbar C}=\frac{N\;\sigma }{\Phi _{0}}$, \ $%
m^{\ast }$ is the effective mass, $\sigma $ the area of the ring.\

$H_{in}=\sum_{j}^{N-1}\frac{\hbar ^{2}}{2\mu
_{j}R^{2}}(-i\frac{\partial }{\partial \varphi
_{j}})^{2}+\sum_{j<i}V_{ji} \ \ \ \ \ \ \ \ \ \  (3)$

\ is the internal Hamiltonian not depending on \ $B$, $\mu _{j}=jm^{\ast
}/(j+1)$ is the reduced mass, and \ the interaction is adjusted as

\qquad \qquad $V_{ji}=\frac{e^{2}}{2\varepsilon
\sqrt{d^{2}+R^{2}\sin^{2}((\theta _{i}-\theta _{j})/2)}}\qquad
\qquad (4)$

 where $\varepsilon $ is the dielectric constant and
the parameter $d$ is introduced to account for the effect of
finite thickness of the ring$^{7}$. \ The Zeeman energy is $\
H_{Zeeman}=\gamma B\cdot S_{Z}\;,$ where $\gamma =g\mu _{B}$, and
$S_{Z}$\ is the Z-component of the total spin $S$.

\ \ \qquad It is clear from (2) that the collective motion is equivalent to
the motion of a single particle with a mass $Nm^{\ast }$ and a charge $Ne$ .
\ However, $H_{in}$ remains to be diagonalized. \ \ Let the eigenstates of $%
H_{in}$ be called internal states, \ the permutation symmetries of their
spatial parts depend on $S$ . \ Thus, it is sure that the internal states
depend on $S$.\ \ On the other hand, the orbital angular momentum $L$ is an
eigenvalue of $H_{coll}$, it describes the collective motion, therefore the
internal states might not depend on $L$.\ \ However, as we shall see,\ they
depend on $L$ in a very special way. \ \ \ In what follows $L$ is considered
as positive, and $B$ is negative.

\qquad\ We shall study mainly the case $N=3$, the basis functions (not yet
normalized)

$e^{i(k_{1}\theta _{1}+k_{2}\theta _{2}+k_{3}\theta
_{3})}=e^{iL\theta _{C}}\cdot e^{i(k_{a}\varphi _{1}+k_{b}\varphi
_{2})}\qquad \qquad (5)$

can be adopted, where \ $L=k_{1}+k_{2}+k_{3}$, $k_{a}=(k_{2}-k_{1})/2$ , and
$k_{b}=k_{3}-L/3$ . \ Since $k_{1},k_{2},$ and $k_{3}$ must be integer to
assure the periodicity , $k_{b}$\ would be an integer if $L=3J$ (where $J$
is an arbitrary integer) and therefore the period of \ $\varphi _{2}$ is $%
2\pi $. Whereas, if $L\neq 3J,$ $k_{b}$\ would not be an integer\ and the
period of \ $\varphi _{2}$ is $6\pi $. \ Obviously, the two kinds of
periodicity of $\varphi _{2}$ lead to two kinds of internal structures. \
Thus, \ together with the two choices $S=1/2$ and 3/2, the internal states
can be classified into four internal series $\psi _{QS,i}$ , where $Q=+1$
implies the case $L=3J$, $Q=-1$ implies $L\neq 3J$.\ \ In other words, $Q$
denotes a specific subset of $L,$ the internal states do not depend on the
value of $L$ but on $Q$.

\qquad\ Due to eq.(1), once $B$ and the energy $E_{QS,i}$ of an
internal state $\psi _{QS,i}$\ is known, a series of energy levels
with angular momenta $L$ belonging to the subset $Q$ can be
generated as

$E=$ $\ D(L+\alpha B)^{2}+E_{QS,i}+\gamma \;B\cdot S_{Z}\qquad
\qquad (6)$

each is associated with the eigenstate

\qquad \qquad $\Psi =\frac{1}{\sqrt{2\pi }}e^{iL\theta _{C}}\;\psi
_{QS,i}\qquad \qquad (7)$

It is noted that $H_{in}$\ does not depend on $B$, therefore $\ E_{QS,i}$
does not depend on $B$ also. \

\qquad\ Let us study the 3-electron ring numerically as an example. \ Let $%
m^{\ast }=0.063m_{e}$, $\varepsilon =12.4$ (for InGaAs), $\;$and$\;d=0.05R$
(the qualitative results are not sensitive to $d$). \ \ For the
diagonalization of \ $H$, the basis functions $e^{i(k_{1}\theta
_{1}+k_{2}\theta _{2}+k_{3}\theta _{3})}/(2\pi )^{3/2}$ are used, \ where $%
k_{i}$\ are integers, each is ranged from -20 to 20. \ This is sufficient to
obtain accurate solutions. \ \ After the diagonalization, the eigenenergies $%
E$\ are obtained. \ From $E$\ and from eq.(6) $E_{QS,i}$\ can be extracted
and four internal series are found just as predicted, they are listed in
Table 1\ .

\vspace{1pt}

\qquad Table 1, \ Internal energies $E_{QS,i}$ of a 3-electron ring (in
meV), the cases with $i=1$ and 2 are listed.

\begin{tabular}{|c|c|c|}
\hline   (Q,S) & $R=30nm$ & $R=90nm$ \\
\hline $(1,\frac{3}{2})$ & 9.1442,\ \  15.1320,$\cdot \cdot \cdot
$
& 2.6533,\ \ 3.6085,$\cdot \cdot \cdot $  \\
\hline $(-1,\frac{3}{2})$ & 11.9455,\ \ 15.4173,$\cdot \cdot \cdot
$
& 3.1107, \ \ 3.6336,$\cdot \cdot \cdot $ \\
\hline $(1,\frac{1}{2})$ & 11.8278,\ \ 11.8827,$\cdot \cdot \cdot
$
& 3.1101,\ \ 3.1105,$\cdot \cdot \cdot $ \\
\hline $(-1,\frac{1}{2})$ & 9.1161, \ \ 11.8567,$\cdot \cdot \cdot
$
& 2.6532,\ \ 3.1103,$\cdot \cdot \cdot $ \\
\hline
\end{tabular}\\

\vspace{1pt}

\qquad It is clear from this table that, not matter how $R$ is, $E_{1,3/2,1}$
and $E_{-1,1/2,1}$are particularly low. \ Using only these two and using
eq.(6) the low-lying spectrum against $B$ can be exactly and entirely
generated\ as shown in Fig.1. \ It is surprising that such a complicated
spectrum contains only two internal states. \ \ Let $L_{o}$ and $S_{o}$
denote the quantum numbers of the ground state. \ In Fig.1 the ground state
band (GSB) contains many segments, each has its own $L_{o}$ and $S_{o}$\ . \
\ When $S_{o}$ =3/2 , $L_{0}=3J$ ; when $S_{o}$ =1/2 , $L_{0}$ $\neq 3J $. \
When $|B|$ increases, $L_{o}$ will increase and the ground state will
transit from one segment to its neighboring segment step by step.

\begin{figure} [htbp]
\centering
\includegraphics[totalheight=3.0in,trim=5 40 5 10]{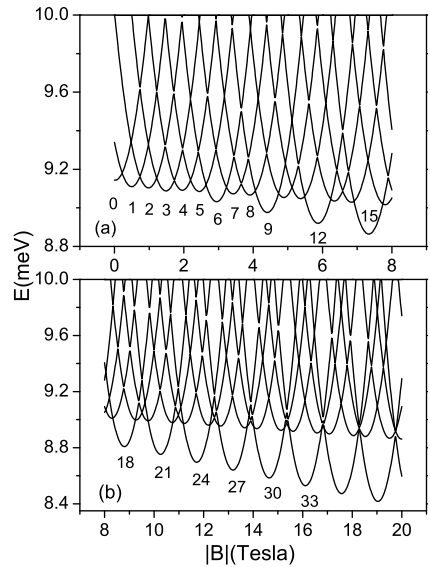}
\caption{Evolution of low-lying levels of a 3-electron ring with\
$R=30nm$ against $B$ . \ The orbital angular momentum of the
ground state$\ L_{o}$\ are marked by the levels.} \label{Fig.1}
\end{figure}

\qquad\ Let $\chi _{\uparrow }(i)$ and $\chi _{\downarrow }(i)$ denote the
up and down spin-states. \ Let $S_{Z}=1/2$ and $\mathbf{\chi }_{\uparrow
\uparrow \downarrow }(123)\equiv \chi _{\uparrow }(1)\chi _{\uparrow
}(2)\chi _{\downarrow }(3),$ then the antisymmetrized internal states can be
expanded as$^{16}$

$\psi _{QS,i}\equiv \psi _{\uparrow \uparrow \downarrow
}(123)\mathbf{\chi }_{\uparrow \uparrow \downarrow }(123)+\psi
_{\uparrow \uparrow \downarrow }(231)\mathbf{\chi }_{\uparrow
\uparrow
\downarrow }(231)+\psi _{\uparrow \uparrow \downarrow }(312)\mathbf{\chi }%
_{\uparrow \uparrow \downarrow }(312)\qquad \qquad \ \ \ \ \ \ \ \
\ \ \ \ \ \ \ \ \ \ \ (8)$

Since the three terms in (8) are different only in the names of particle,
the observation of one of them is sufficient. \ The norm $|\psi _{\uparrow
\uparrow \downarrow }(123)|$ of $\psi _{QS,i}$ with $i=1$ and 2 are plotted
in Fig.2 as a function of $\varphi _{2}$. \ In this choice, we are observing
the motion of the spin-down electron relative to the two spin-up electrons.
\ It is reminded that, if a pair of electrons are close to each other, the
energy would increase due to the $e-e$ repulsion. \ Thus, the best
configuration is the equilateral triangle ($ET$) which is associated with
the minimum of the total interaction energy. \ From the point of dynamics,
all the first-states ( $i=1$) would pursue the $ET$ so as to lower their
energies. \ However, only the two series with $(QS)=(1,\frac{3}{2})$ and (-1,%
$\frac{1}{2})$ succeed, they have a peak at $\varphi _{2}=\pi $ which
corresponds to an $ET\;$(Fig.2a and 2d). \ Whereas there is a node at the $%
ET $ in the other two series . \ Evidently, the node appearing at the
minimum of potential energy would push the wave function away from the
minimum and thereby cause the excitation of an oscillation \vspace{1pt}back
and forth around the minimum as an equilibrium point, this would cause a
great increase in energy. \ This explains why the other two series are
remarkably higher.

\qquad The appearance of the node has a profound background of
symmetry. \ \ In general,  if the $N$ electrons form a N-sided
equilateral polygon ($EP$)in the ring, the total interaction
energy would be minimized. \ However, at this configuration, a
rotation by $2\pi /N$\ is equivalent to a cyclic permutation of
coordinates. \ It was proved in ref.\lbrack 15\rbrack\ that this
equivalence leads to a constraint. \ As a result, the wave
functions of $S=N/2$ states would have an inherent node at the
$EP$ if $L\neq N\;G$ , where $G$\ is an integer (half-integer) if
$N$\ is odd (even). \ Alternatively, when $S=N/2-1,$ the node
would appear if $L=N\;G,$\ \ . \ This is called an
$EP-$prohibition (PEP). \ Obviously, if the set of orbital angular
momenta\ contained in an internal series are free from the PEP,
the series is allowed to pursue the favourable $EP$ geometry,
therefore its first states ($i=1$) would be particularly low. \
This is the symmetry background that, when $N=3$\ , the
($Q,S$)=(1,3/2) and (-1,1/2) first-states are low (cf. Table 1).

\qquad\ Based on eq.(6), the feature of the spectra can be
explained analytically. \ When $|B|$ is small, Fig.1 shows that,
in the GSB, the difference in $L_{o}$ of two neighboring segments
is one, i.e., $L_{o}$ is compactly aligned. \ Each segment is
ranged from $B_{a}(L_{o})$\ to $B_{b}(L_{o})$ and contains a
minimum of energy at  $ B_{\min }(L_{o})$. $B_{a}$\ and $B_{b}$
can be derived from eq.(6) together with the condition that
neighboring segments have the same energy at the boundaries.
Besides, $ B_{\min }$ can also be derived from eq.(6) and from
$dE/dB=0$. Due to $\gamma $ being small, we have

$B_{\min }(L_{o})=-(2D\alpha L_{o}+\gamma S_{o})/2D\alpha
^{2}\approx -L_{o}/\alpha , \ \ \ (9)$

\ Accordingly, \ the distance between the minima of two
neighboring segments is $\ B_{\min }(L_{o})-B_{\min }(L_{o}+1)$ $\approx 1/\alpha =$ $%
\Phi _{o}/3\sigma $\ , which is just the period of the FABO$^{2,5}$. \ \
Thus, the compact alignment \ leads to the FABO. \

\begin{figure} [htbp]
\centering
\includegraphics[totalheight=2.6in,trim=30 30 30 30]{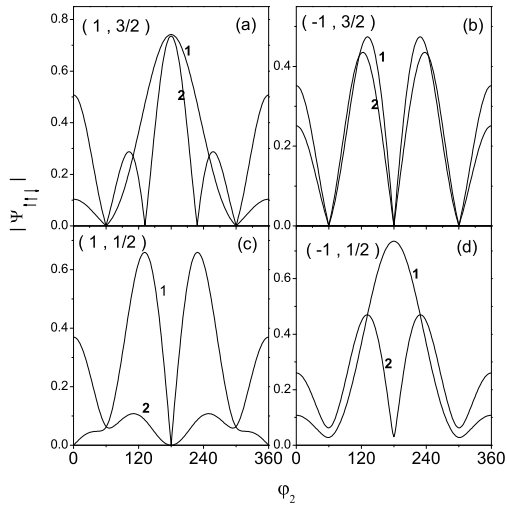}
\caption{  $|\psi _{\uparrow \uparrow \downarrow }|$ component\
of\ the internal states $\psi _{QS,i}$ plotted against $\varphi
_{2}$ , $\varphi _{1}$ is fixed \ at 120$^{\circ }$, $R=30nm$, the
label $(QS)$\ are marked in the upper corner and $i$\ \ by the
curves.} \label{Fig.2}
\end{figure}

\qquad\ Fig.1 shows that , due to the Zeeman effect, the increase
of $|B|$ would enlarge the segments with $S=3/2$ (i.e., $L_{o}=3J$
), while those with $S=1/2$ (i.e., $L_{o}\neq 3J$ ) \ would become
smaller and smaller. \ When $|B|$ is sufficiently large, the
minima of the $S=1/2$ segments would disappear. \ This would begin
to occur if\ $B_{b}(3J)=B_{\min }(3J+1)$. \ In this case we define
a critical value $B_{crit1}=|B_{a}(3J)|.$ \ When $|B|$ is even
larger, all the $S=1/2$ segments would disappear. \ This would
begin to occur if\ $B_{b}(3J^{\;\prime })=B_{a}(3J^{\;\prime
}+3)$. \ In this case we define another critical value
$B_{crit2}=|B_{a}(3J^{\;\prime })|.$ \ E.g., for
Fig.1, $B_{crit1}=9.8$\ and $B_{crit2}=18.2.$ \ \ Evidently, \ when \ $%
|B|>B_{crit1}$, \ due to the miss of the minima of the $S=1/2$ segments, the
distance between two adjacent minima (evaluated from (9)) is changed from $%
\Phi _{o}/3\sigma $\ to $\Phi _{o}/\sigma .$ \ Thus, the period of
the normal ABO recovers, and $B_{crit1}$ is the boundary
separating the fractional and normal ABO. \ When \
$|B|>B_{crit2}$, all the $S=1/2$
segments disappear , and the GSB\ becomes purely polarized. \ Thus $%
B_{crit2}$ is the boundary separating the partial and full
polarization.

\begin{figure} [htbp]
\centering
\includegraphics[totalheight=1.6in,trim=20 40 30 30]{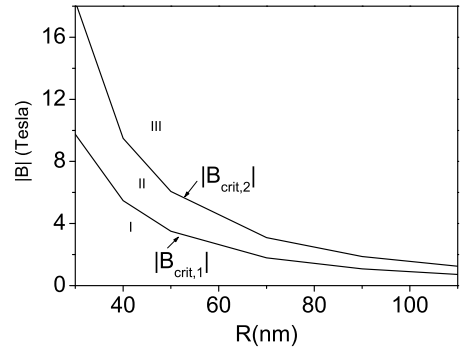}
\caption{ The evolution of $B_{crit1}$ and $B_{crit2}$ against
$R$.} \label{Fig.3}
\end{figure}

Based on the analytical expression eq.(6), the values of
$B_{crit1}$ and $B_{crit2}$ can be obtained analytically, they
depend on the radius $R..$ \ The numerical results are plotted in
Fig.3, which is a phase diagram\ containing three regions. \ The
FABO appears only in region (I), thus it emerges only when $|B|$\
is weak. \ However, for the rings with a small $R$ (say, $R<30
nm$), the FABO\ exists even if $B$ is strong. \

\vspace{1pt}\qquad Incidentally, let the orbital angular momenta of the GSB
associated with $B_{crit1}$ and $B_{crit2}$ be denoted as $L_{crit,1}$ and $%
L_{crit,2}$. \ \ It was found that both $L_{crit,1}$ and $L_{crit,2}$ depend
on $R$ very weakly. \ When $R$ is ranged from 30 and 120, $L_{crit,1}$\ and $%
L_{crit,2}$ remain to be 21 and $36,$ respectively.

\qquad The transitions of $L_{o}$\ and $S_{o}$ in the GSB would lead not
only to the oscillation of ground state energy, but also an oscillation in
dipole radiation. \ Let the temperature $T$ be very low (e.g., $T\leq 0.01K$%
) so that the system is mainly in its ground state. \ When dipole
transitions between the ground state and a final state occurs both $S_{o}$
and its Z-component are conserved, and the difference in angular momenta $%
L_{f}-L_{o}$ is $\ \pm 1$, ( the subscript $f$\ denotes the final state). \
The difference in energies, from eq.(6), is

$E_{f}-E_{o}=E_{Q_{f},S_{o},i}-E_{Q_{o}S_{o},1}+D(1\pm
2(L_{o}+\alpha B)) \ \ (10)$

Let us focus on the most infrared radiation, i.e., the low energy limit
denoted as $\lim (E_{f}-E_{o})$ . \ When $L_{o}=3J,$ we have $%
(Q_{o},S_{o})=(1,3/2)$. \ Since meanwhile $L_{f}=3J\pm 1$, thus
$Q_{f}=-1$ . \ Since $E_{-1,3/2,1}$ is associated with the
internal \ series suffering the PEP, it is high and therefore
leads to a big $\lim (E_{f}-E_{o})$ (cf. Table 1). \ When
$L_{o}=3J-1,$ we have $(Q_{o},S_{o})=(-1,1/2)$. \ Meanwhile
$L_{f}$ has two choices, $3J-2$ or $3J$. \ The former choice has
$Q_{f}=-1$, thus the ground state and the final state would have
the same internal state $\psi _{-1,\frac{1}{2},1}$\ with the same
internal energy $E_{-1,1/2,1}$, and
therefore leads to a small $\lim (E_{f}-E_{o}).$ \ The latter choice has $%
Q_{f}=+1$. \ However, $E_{1,1/2,1}$ is \ also pushed up by the PEP,
therefore this choice leads to a large $E_{f}-E_{o}$ and can be neglected. \
\ Similarly, When $L_{o}=3J+1,$ a small $\lim (E_{f}-E_{o})$ exists. \ The\
low energy limit of the radiation absorbed by the GSB is plotted in Fig.4. \
\ During the\ increase of $|B|<B_{crit1}$, each time when $B$\ lies in a
segment with $L_{o}=3J,$ the limit $\lim (E_{f}-E_{o})$\ jumps up remarkably
and then jumps down afterward.  This causes a very strong oscillation much
stronger than the oscillation of the ground state energy. \ This oscillation
would become even stronger when $R$\ becomes smaller. The period of the
oscillation is simply $B_{\min }(L_{o})-B_{\min }(L_{o}+3)$ $\approx $ $\Phi
_{o}/\sigma $, the same as the normal ABO. \ However, when $%
|B|>B_{crit1},\;$the GSB\ is dominated by $L_{o}=3J$\ states, and the
oscillation would disappear because the $\lim (E_{f}-E_{o})$ do not jump
down again. \

\begin{figure} [htbp]
\centering
\includegraphics[totalheight=2.2in,trim=50 80 20 60]{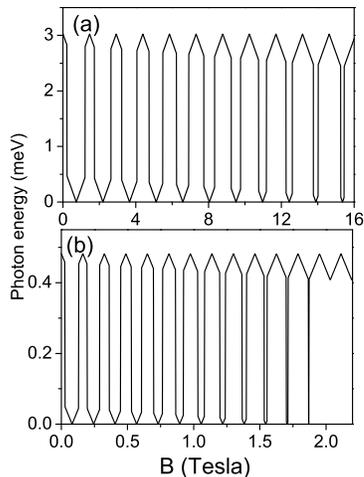}
\caption{Oscillation of the low energy limit of the radiation of
the ground state against $B$. \ $R=30nm$ (a) and $90nm$ (b).}
\label{Fig.4}
\end{figure}

\qquad In summary, the narrow rings with three electrons have been studied.
\ The following findings are noticeable.

\qquad (1) When the relative angles $\varphi _{i}$ are introduced to
describe the internal states, the periods of $\varphi _{i}$ depend on $L.$ \
Therefore, the internal states depend on $L$ in a specific way, and can be
classified into a few series based on the periodicity. \ A specific subset
of $L$\ is allowed in a series.

\qquad (2) \ The structures of distinct series of internal states are
affected by symmetry in different ways. \ In particular, two favourable
series are free from the PEP, their first-states \ are particularly low. \ \
These two first-states are sufficient to generate exactly the complete
low-lying spectra. \ This is an interesting point, thereby the spectra can
be understood analytically.

\qquad (3)  The $L$ allowed \ by the two first-states are complementary, \
they cover the whole range of $L$ . When $|B|$\ increases but $\leq B_{crit1}
$, the two first-states appear in the GSB\ alternatively. \ As a result, \ $%
L_{o}$ increases one-by-one, i.e., they are compactly aligned.  This leads
to the FABO. \ However, when $B>B_{crit1},$\ the internal state with $%
S=1/2$\ disappears from the GSB. \ As a result, \ $L_{o}$\ increases each
time by three, thus the normal ABO recovers.

\qquad (4) When $|B|<B_{crit1}$, a strong oscillation of the low
energy limit of the radiation of the GSB\ was found. \ The
amplitude of this oscillation is much stronger than that of the
ground state energy. This is also a very interesting phenomenon.

\ \ \ \ Although the above findings are extracted from 3-electron
rings, they hold also for $N>3$ rings. \ It implies that the
internal states can also be classified according to the periods of
$\varphi _{i}$, thus they depend on $L$ in a special way.  Some
internal series are free from the PEP, and the low-lying spectra
are generated by the first-states of them. When $|B|$\ is small,
$L_{o}$ is compactly aligned, the FABO would emerge. When $|B|$\
is large ($|B|>B_{crit2}$), the GSB is fully polarized and $L_{o}$
increases each time by $N$, thus the normal ABO recovers. Before
the full polarization , there would be a domain of $B$ in which
the GSB is dominated by both $S=N/2$ and $N/2-1$ segments. When
$B$ is given in one of the $S=N/2$ segment, the most infrared
radiation of the ground state is associated with a transition of
the internal state from a PEP-free series to a series suffering
the PEP, and thus results in a large $lim(E_{f}-E_{o})$, while in
the neighboring segments the $lim(E_{f}-E_{o})$ is small. This
leads to the occurrence of the above mentioned optical
oscillation. Incidentally, this oscillation does not appear when
$N=1$ or 2.

Nonetheless, the cases with $N>3$ and the cases of broad rings
have to be further studied.

\vspace{1pt}

Acknowledgment, \ This work is supported by the NSFC of China under the
grants 90306016  and 10174098.

\end{document}